# Single Image Dehazing Algorithm Based on Sky Region Segmentation

LI Weixiang , JIE Wei, S. MahmoudZadeh

[1,2] Nanjing Tech University, Nanjing Jiangsu 211800, China
[3] Deakin University Geelong, Victoria 3220, Australia
lwxlf@njtech.edu.cn

**Abstract.** In this paper a hybrid image defogging approach based on region segmentation is proposed to address the dark channel priori algorithm's shortcomings in de-fogging the sky regions. The preliminary stage of the proposed approach focuses on segmentation of sky and non-sky regions in a foggy image taking the advantageous of Meanshift and edge detection with embedded confidence. In the second stage, an improved dark channel priori algorithm is employed to defog the non-sky region. Ultimately, the sky area is processed by DehazeNet algorithm, which relies on deep learning Convolutional Neural Networks. The simulation results show that the proposed hybrid approach in this research addresses the problem of color distortion associated with sky regions in foggy images. The approach greatly improves the image quality indices including entropy information, visibility ratio of the edges, average gradient, and the saturation percentage with a very fast computation time, which is a good indication of the excellent performance of this model.

## 1 Introduction

One of the main challenges associated with the outdoor computer vision system in foggy weather is that the captured images show degradation phenomena such as decreased contrast and color distortion [1]. Therefore, having an efficient de-hazing mechanism is a necessary requirement to cope with the raised issues in dimmed or foggy weather conditions [2]. There are two main approaches for image defogging, one is relying on image enhancement methods [3], such as Retinex algorithm [4], wavelet transform [5], etc., and another approach is the image restoration. Image enhancement techniques can effectively improve the contrast of foggy image and boost the visual effect; however, some image information will be lost as the image degradation with fog is not taken into account in these methods.

On the other hand, the image restoration method uses the modeling and analysis of atmospheric scattering to establish a foggy image degradation model, and takes advantage of the prior knowledge captured in fog-free weather to recover the degraded foggy image [6]. Respectively, Tan (2008) applied contrast enhancement to restore the structural features of images and improve the Visibility in bad weather condition [7]. In [8], reflectivity of the object surface is assumed to be not correlated with the medium transmission rate. The author applied independent component analysis to obtain the scene reflectivity, while Markov Random Field (MRF) is employed for image restoration. However, for regions with low image signal-noise, the applied method in [8] is not efficient in terms of defogging. Later on, He et al., [9] proposed a dark channel prior image defrosting algorithm, which could achieve reasonable performance in enhancing majority of the foggy images. However, refining the transmission rate in this method relies on soft matting algorithm, which results in excessive time complexity.

The shortcoming with the dark channel prior theory is that applying this method on images with sky region results in color distortion while reducing the defogging effect, which makes it inappropriate for the images with the sky region. To address this issue, a guided filtering method has been applied by [10] to improve the efficiency of dark channel prior fogging and reduce the time complexity associated with soft matting method; however, in some cases the method tends leave defogging phenomenon incomplete or patchy. Liu et al., [11] proposed a fast image defogging algorithm, which could also improve the sharpness and contrast of the restored image, but this method is still not ideal for the defogging effect of the sky region. Later on an integrated pipeline has been introduced by Li et al., in (2017) that could directly reconstruct a fog-free image through an end-to-end Convolutional Neural Networks (CNNs), but still the resultant image tends to be dark and not clear enough [12].

Respectively, a single image defogging algorithm based on region segmentation is proposed in this research, which aims to address the shortcomings associated with the aforementioned defogging algorithms. To this purpose, MeanShift algorithm with embedded confidence is applied in the first stage to segment the sky and non-sky regions of the original image. Afterward an optimized dark channel prior defogging algorithm is adopted to handle defogging of the non-sky area, while the sky region defogging is handled using an improved neural network DehazeNet. Using such a hybrid approach effectively solves the problem of sky region color distortion in the dark channel prior algorithm. Performance of the proposed approach is investigated and compared to some of the previous researches in the area. The captured restored images in this research tend to have supreme contrast and excellent brightness, rich details, where the resultant colors are very natural, which is a good indication of the promising performance of this hybrid approach.

## 2    Dark Channel Prior Defogging Algorithm

### 2.1    Atmospheric Scattering Model

The quality reduction process of foggy image is generally represented by the following atmospheric scattering model:

$$I(x) = J(x)t(x) + A(1-t(x))  \qquad (1)$$

where, $I(x)$ is the observed hazy image, $J(x)$ is the real scene to be recovered, $A$ is the global atmospheric light, $t(x)$ is the medium transmission.

### 2.2 Dark Channel Prior Defogging Principle

He, et al., [9] found that in the vast majority of non-sky areas, the color channel brightness value of some pixels is close to 0. For any input image $J$, Its dark channel $J^{dark}$ satisfies:

$$J^{dark}(x) = \min_{y \in \Omega(x)} \left( \min_{c \in \{r,g,b\}} (J^c(y)) \right) \to 0 \tag{2}$$

In (2), $\Omega(x)$ corresponds to local neighborhood centered on $x$; $J^c$ denotes an RGB channel of the image $J$. Accordingly, the given equation in (1) can be transformed to the following:

$$\frac{I^c(x)}{A^c} = t(x)\frac{J^c(x)}{A^c} + 1 - t(x) \tag{3}$$

Assuming the transmission rate as a constant value in the selected window (presented by $\tilde{t}(x)$ here) and applying the *min* function, the given relation in (3) will be turned to the following:

$$\min_{y \in \Omega(x)} \left( \min_{c \in \{r,g,b\}} \frac{I^c(y)}{A^c} \right) = \tilde{t}(x) \min_{y \in \Omega(x)} \left( \min_{c \in \{r,g,b\}} \frac{J^c(y)}{A^c} \right) + 1 - \tilde{t}(x) \tag{4}$$

Given the dark channel prior theory:

$$\min_{y \in \Omega(x)} \left( \min_{c \in \{r,g,b\}} \frac{J^c(y)}{A^c} \right) = 0 \tag{5}$$

Substituting into equation (4):

$$\tilde{t}(x) = 1 - \min_{y \in \Omega(x)} \left( \min_{c \in \{r,g,b\}} \frac{I^c(y)}{A^c} \right) \tag{6}$$

In order to express the depth of field for the image, the coefficient $\omega$ ($0 < \omega < 1$) is introduced by the equation (6) as follows:

$$\tilde{t}(x) = 1 - \omega \min_{y \in \Omega(x)} \left( \min_{c \in \{r,g,b\}} \frac{I^c(y)}{A^c} \right) \tag{7}$$

Given $A$, $t$ and $I$, we can obtain $J$ as follows:

$$J(x) = \frac{I(x) - A}{\tilde{t}(x)} + A \tag{8}$$

According to (8), the $\tilde{t}(x)$ has inverse relation to $J$, respectively, when the value of transmission $\tilde{t}(x)$ is too small, $J$ will be large, which leads to the overall whitening of the restored image. Therefore, the threshold value of $t_0$ (usually is set to 0.1) is encountered to obtain the final image restoration formula as follows:

$$J(x) = \frac{I(x) - A}{\max(\tilde{t}(x), t_0)} + A \tag{9}$$

## 3 Research Methodology

In this research, we employed a hybrid method including three different phases. In the first stage before defogging procedure, the image gets segmented into sky and non-sky regions. In the second stage, the non-sky regions will be defogged using an improved dark channel prior defogging algorithm. Afterward, the sky region gets proceeded using deep learning neural network DehazeNet. Ultimately, the reconstructed image is obtained by combining the processing results of sky and non-sky regions.

### 3.1 Extract and Segment the Sky Region

This research, sky region segmentation is carried out using an edge detection embedded with confidence measurement strategy along with the mean shift approach. The confidence measurement strategy facilitates more efficient edge detection.

Assuming the $x_i$, $z_i$, ($i=1,2,…,n$) respectively represent the d-dimensional input image of the spatial-color domains, which at the same time corresponds to the pixel of the filtered image, and $L_i$ denote the label of the $i^{th}$ pixel of the segmented image, the mean shift segmentation algorithm consists of the following steps:

***Step*-1:** Let $j=1$, $y_{i,1}=x_i$

***Step*-2:** Use the following formula to calculate $y_{i,\,j+1}$, until the convergence of $y=y_{i,c}$.

$$y_{j+1} = \frac{\sum_{i=1}^{n} x_i g(\|\frac{x-x_i}{h}\|^2)}{\sum_{i=1}^{n} g(\|\frac{x-x_i}{h}\|^2)}, j=1,2,\cdots \quad (10)$$

Where, $y_1$ is the initial center of the core; $y_{j+1}$ is the weighted average value of position $\{y_j\}$, ($j=1,2,…,C$) calculated by Gaussian kernel function.

***Step*-3:** Let $z_i = (\mathrm{x}_i^s, \mathrm{y}_{i,c}^r)$, where $S$ represents the domain of the eigenvector and $r$ is the color domain information.

***Step*-4:** For each $z_i$ in *Step*-3, put the points whose distance with $z_i$ is less than $h_r$ in the color domain, and less than $h_s$ in the airspace domain, into the corresponding category.

***Step*-5:** Set the threshold value $M$ to exclude the number of pixels smaller than $M$.

The whole procedure is depicted by Fig.1.

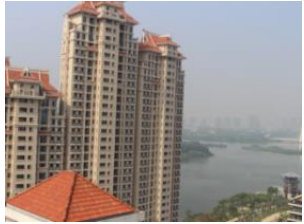 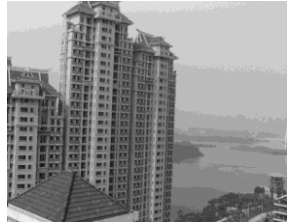

(a) The original image      (b) Grayscale image

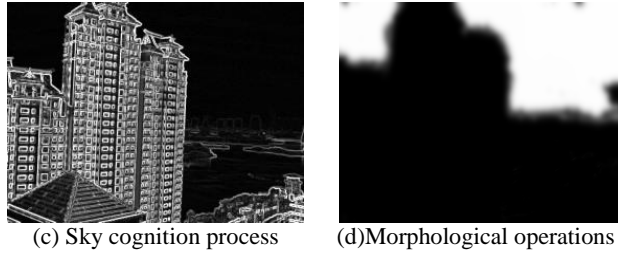

(c) Sky cognition process    (d) Morphological operations

**Fig. 1.** Sky segmentation mechanism

As shown in Fig.1, first the original foggy image is transformed to grayscale image (Fig.1(b)). The introduced method is then applied to segment different regions of the image (given by Fig.1(c)). Afterward, morphological expansion along with corrosion operations is conducted to extract the sky area and obtain the binary image given in Fig.1(d).

### 3.2 Dark Channel Prior Optimization Algorithm

**Acquiring the Atmospheric Light A.** Most of the dark channel prior algorithms take the brightest pixels in the image with fog as the value of atmospheric light A. The main issue with this strategy is that if the non-sky region includes a bright area as well, the obtained atmospheric light A will be larger, which will affect the defogging efficiency. To deal with this issue, we need to scan the whole image and take the average of the pixel point's value with their brightness of the first 0.5%, and then assign it to atmospheric light value of A. The atmospheric light obtained by this method has strong robustness and can eliminate noise interference to some extent.

**Obtaining the Transmitter t.** In this research, the value of dark channel $J^{dark}$ is assumed to be maximized in order to reduce the transmitter t to a particular extent. Therefore, the min operation in (2) is replaced with the average operation as follows:

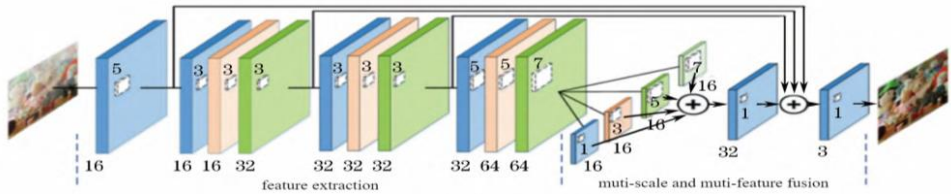

**Fig. 2.** Improved DehazeNet structure

$$J^{dark}(x) = \underset{y \in \Omega(x)}{avg} \left( \underset{c \in \{r,g,b\}}{\min} (J^c(y)) \right) \quad (11)$$

According to equations (1) and (7), the improved transmitter can be obtained by (12) as follows:

$$\tilde{t}(x) = 1 - \omega \, avg_{y \in \Omega(x)} \left( \min_{c \in \{r,g,b\}} \frac{I^c(y)}{A^c} \right) \tag{12}$$

Finally, guided filtering is used to refine the transmitter.

**Brightness Adjustment.** The transmitter obtained by dark channel prior algorithm tends to be small, so this cause the resultant defogged image to be dark. Therefore, in order to compensate this issue, current research conducts a nonlinear contrast stretch image enhancement method [13] to adjust the brightness of the restored image, which is formulated as follows:

$$J'(x) = J(x)(2 - \frac{mean(J(x))}{255}) \tag{13}$$

Where, *mean* ($J(x)$) represents the pixel brightness. The given adjustment through the (13) can improve the total brightness of the restored image.

### 3.3 Improved DehazeNet Defogging Algorithm

In order to have an efficient defogging, Cai et al., [14] proposed an end-to-end system model DehazeNet based on CNN, which is divided into four stages: feature extraction, multi-scale mapping, local extremum and nonlinear regression.

This research applies the optimized version of the DehazeNet model making the use of full convolution network structure, which effectively preserves the important image information. Consequently, the method is augmented with a multiple feature fusion mechanism facilitating the direct reconstruction of the defogged sky-region image out of the foggy image. So the algorithm follows two steps of multi-scale feature fusion and feature extraction. The operation diagram of the algorithm is depicted by Fig.2.

**Feature Extraction.** Feature extraction stage consists of a single convolutional layer and three consecutive three-layer convolutional layers. The output of all convolution operations should be equal to zero to ensure that the output image remains in the same size. The given 9 convolutional layers operate as 3 sets of feature extraction units which can gradually obtain the detailed information of the image. The convolutional operation is formulated as follows:

$$G_l = W_l * F_{(l-i)}(Y) \tag{14}$$

Here, $W_l$ represents the $l^{th}$ convolutional layer and corresponds to the number of parameters calculated through the $n_l \times f_l \times f_l$, where $n_l$ and $f_l$ are the number and size of the convolution kernel, respectively. $G_l$ is the characteristics of the output image, and '*' is the convolution operation.

The PReLU is adopted as the network's activation function in this research. The PReLU function is advantageous to be able to modify the parameters, which improves the model's overall scalability and generalization. Furthermore, PReLU can accelerate the convergence speed of the algorithm. The PReLU function is defined as follows:

$$P_{\text{Re}LU}(x_i) = \max(x_i, 0) + a_i \min(0, x_i) \tag{15}$$

Where, $x_i$ is the positive input signal to layer $i$; $a_i$ is the weight coefficient of the negative interval of the $i^{th}$ layer. Accordingly, output of the final convolutional layer would be:

$$F_l(Y) = P_{ReLU}[W_l * F_{l-1}(Y) + B_l] \qquad (16)$$

Here, $B_l$ represents the bias to layer $l$, and $F_l$ is the final characteristics of the output graph.

**Multi-scale Multi-feature Fusion.** Since the image of foggy sky contains both color differences caused by different fog concentrations and texture features, this paper adopts multi-scale multi-feature fusion method to further extract the existing features. The model is structured with five trainable convolutional layers, where the output of these multi-scale convolutional layers is fed to three feature extraction units as input. Multi-feature fusion is completed with four cores of 1×1, 3×3, 5×5, and 7×7. Multi-scale multi-feature fusion layer parameters are configured as is shown in Table 1.

**Table 1.** Multi-scale Multi-feature fusion parameter configuration

| Layer type | Configuration(padding, Stride, feature map, Kernel) | | | | |
|---|---|---|---|---|---|
| Convolution layer | Pad-0 | Stride-1 | Fm-16 | Kernel-1×1 | PReLU |
| Convolution layer | Pad-0 | Stride-1 | Fm-16 | Kernel-3×3 | PReLU |
| Convolution layer | Pad-0 | Stride-1 | Fm-16 | Kernel-5×5 | PReLU |
| Convolution layer | Pad-0 | Stride-1 | Fm-16 | Kernel-7×7 | PReLU |

The multi-scale convolution process is as follows:

$$F_{li}(Y) = P_{ReLU}[W_{li} * F_{l-1}(Y) + B_l] \qquad (17)$$

In (17), $B_l$ represents the bias to layer $l$, $F_{l-1}$ is the characteristics of the output graph, $W_{li}$ ($i$=1, 2, 3, 4) corresponds to four convolution kernels. Four given sets of convolution kernels process in parallel in order to provide 64 feature graphs. Afterward, the provided 64 feature graphs will be convolved with a convolution kernel size of 1×1. After all, multi-scale feature information gets weighted and fused to obtain the image after defrosting.

## 4  Simulation Results and Analysis

A number of foggy images are selected for simulation and evaluation of the applied method. The efficiency of the proposed method in this paper is compared to the algorithms applied in the same area by two recent approaches [9], [14]. The provided results and the comparisons are depicted in Fig.3, which is carried out to test and evaluate the applied methods' performance from the various perspective and in different environments. The mainstream deep learning caffe and Matlab R2014a have been used as the simulation platform. The PC used for the experiment was configured as: 64-bit Windows10 operating system, 8GB of memory, Intel Core(TM)i7-6700K@4.0GHz processor, and NVIDIA QUADRO GP100 GPU.

For training and testing the applied improved DehazeNet model, an experimental data source of 10 foggy images has been used, in which the images are captured under natural scenes and 60 images of foggy images restored by atmospheric scattering model. The images have been rotated for 90°, 180°, and 270°, and then expanded to 2, 3, 4, and 5 times to get 1400 pictures of different resolution and scene. The provided images then have been used as the training and validation sets with of 6:1, respectively.

### 4.1  Subjective Evaluation

The efficiency of the proposed approaches by He et al., [9] and Cai et al., [14] in defrosting the image of non-sky areas is relatively good as the image clarity and contrast are greatly improved, which is presented by Fig.3. However, referring to the results depicted in Fig.3(b)(c) and comparing them to Fig.3(a)(d), apparently both of the proposed approaches in [9] and [14] do not show the expected efficacious performance in dealing with sky regions. Especially focusing on Fig.3 (b) and Fig.7(b) it is obvious that although the proposed approach in [9] greatly defogs the non-sky areas, but color distortion occurs in non-sky regions. Cai et al., [14] took the advantages of end-to-end CNN model to deal with this issue; however, the method provided poor defogging performance in the natural scene, which is clearly presented by Fig.4.

Furthermore, referring to Fig.3 to Fig.7 it can be seen the defogged images by [9] and [14] tend to be dark with low contrast, which negatively affect the visual appearance of the scene. Comparing to the proposed methods by [9] and [14], the introduced hybrid method in this research shows competent performance in defogging the image of both sky and non-sky area, while also improves the overall brightness of the restored image, making the color of the image real and natural. Performance of the proposed hybrid method and the previously applied methods is depicted by Fig.3 to Fig.7 and the comparison is carried out in different scenes.

Considering the both sky and non-sky regions in Fig.3 and Fig.5, evidently our proposed approach greatly preserves the images natural of the original scene in the close shot when comparing to [9] and [14]. Moreover, defogging effect at the distant shot is greatly improved, and the method tends to show a superior performance in the overall color reduction. Accordingly, the provided comparison and analyses by Fig.3 to Fig.7 is a good indication of superior performance of the proposed model in this research.

| (a) | (b) | (c) | (d) |
|---|---|---|---|
| The original image | Algorithm applied by He et al., [9] | Algorithm applied by Cai et al., [14] | Algorithm proposed by this research |

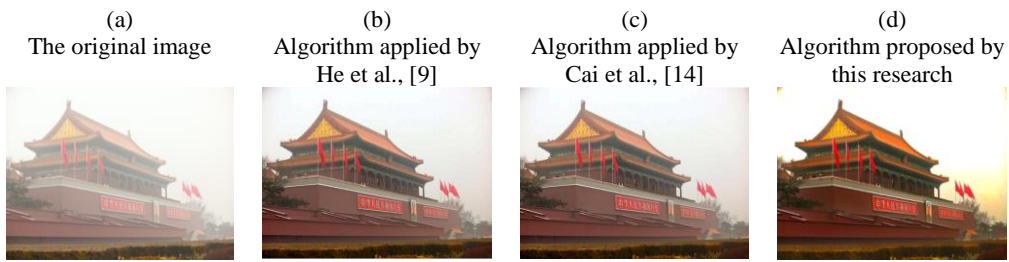

**Fig. 3.** Comparison of defogging effect in sky/non-sky regions

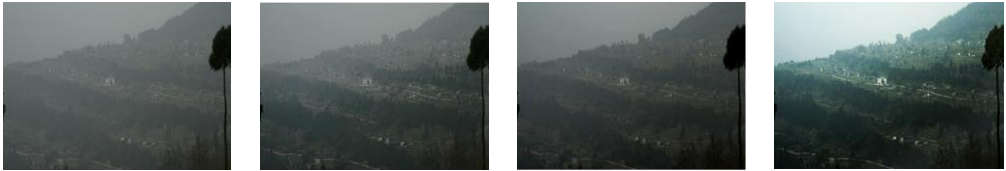

**Fig. 4.** Comparison of defogging effect in natural Scene

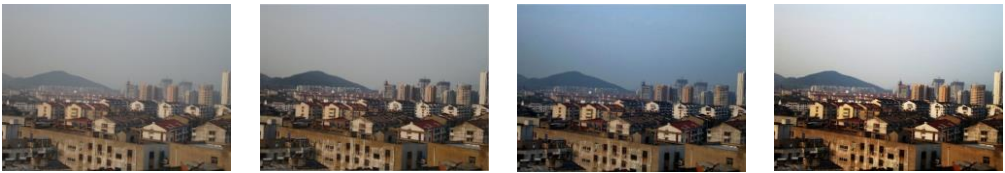

**Fig. 5.** Comparison of defogging effect in building Scene

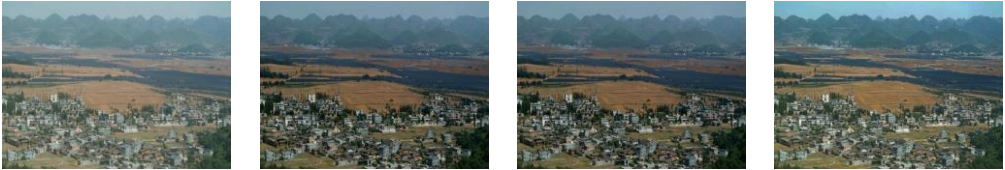

**Fig. 6.** Comparison of defogging effect in natural Scene

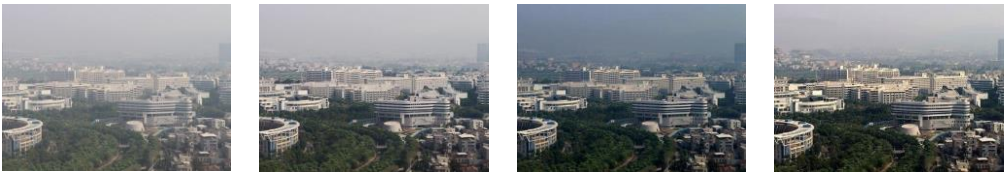

**Fig. 7.** Comparison of defogging effect in city Scene

### 4.2 Objective Evaluation

To further assess the performance of our model, number of important image quality evaluation indices such as entropy information, visibility ratio, average gradient, and pixels' saturation percentage are investigated and compered to [9] and [14].

This investigation and comparison is provided by Table.2, where the entropy information represents the richness of information contained in an image, which should be maximized. Both ratio of new visible edges and the percentage of saturated pixels reflect the degree of local contrast enhancement of foggy images. The ratio of new visible edges needs to be increased in order to improve the clarity of the image and feature recovery of the scene. On the other hand, the saturated pixels' percentage should be reduced in order to achieve a better image restoration.

The average gradient reflects the details of the image, which has a direct relation to image contour clarity and should be increased [15].

**Table 2.** Comparison of performance indices provided by different algorithms

| Image | Applied Method | Entropy Info | Visibility Ratio | Average Gradient | Pixels' Saturation % | Run Time (s) |
|---|---|---|---|---|---|---|
| A<br>473×283 | Cai et al., [14] | 5.7105 | 0.0276 | 3.8103 | 15.0008 % | 11.7350 (s) |
| | He et al., [9] | 5.7130 | 0.1146 | 3.6574 | 32.2857 % | 3.14680 (s) |
| | This Research | 5.7706 | 0.2492 | 4.0763 | 12.9594 % | 2.19860 (s) |
| B<br>472×317 | Cai et al., [14] | 6.8610 | 0.2695 | 3.6932 | 18.7260 % | 13.3360 (s) |
| | He et al., [9] | 6.8488 | 0.1388 | 3.1168 | 16.5730 % | 4.19320 (s) |
| | This Research | 6.8989 | 0.3346 | 3.7371 | 13.8560 % | 3.30620 (s) |
| C<br>470×352 | Cai et al., [14] | 5.4778 | 0.0710 | 3.2199 | 20.3762 % | 9.53600 (s) |
| | He et al., [9] | 7.4709 | 0.1475 | 2.9654 | 18.2453 % | 2.74060 (s) |
| | This Research | 7.5389 | 0.1642 | 3.6061 | 17.6342 % | 2.13450 (s) |
| D<br>600×525 | Cai et al., [14] | 6.4162 | 0.2318 | 5.7956 | 15.6282 % | 12.2658 (s) |
| | He et al., [9] | 7.3935 | 0.3237 | 5.7364 | 15.6228 % | 1.83980 (s) |
| | This Research | 8.4941 | 0.3020 | 5.8585 | 14.3758 % | 2.16740 (s) |
| E<br>564×399 | Cai et al., [14] | 7.5322 | 0.1983 | 1.9538 | 15.9823 % | 12.2486 (s) |
| | He et al., [9] | 8.3514 | 0.0584 | 1.7980 | 14.1240 % | 2.23650 (s) |
| | This Research | 8.4609 | 0.1874 | 2.2166 | 14.0026 % | 3.07550 (s) |

Given the results in Table. 2, it can be seen the entropy information and average gradient of the algorithm in this research tends to be higher that provided values by [9] and [14], which indicates superior color saturation and methods accuracy in detailed image restoration. Fast computation is also an important performance factor that should be taken into account. In terms of algorithm execution time, our propose approach considerably reduces the run time in the two stages of sky region segmentation and defogging comparing to [9] and [14], which also indicates method's competent execution efficiency.

The simulation results show that the proposed algorithm is thoroughly suitable for non-sky region defogging as well. The proposed method in this research provides great level of color restoration, saturation and clarity improvement compared to [9] and [14]. From the objective perspective, the performance indices obtained by the algorithm in this paper are also superior to the results provided by [9] and [14]. Ultimately, it can be inferred from the entire comparisons over the simulation results that our proposed method tends to have excellent operational efficiency and great performance especially in processing of the single foggy image.

## 5   Conclusion

In this paper, a single image defogging algorithm based on region segmentation is proposed, which uses the optimized atmospheric dissipation function to defog the sky region. The proposed approach employs DehazeNet algorithm which is based on deep learning convolutional neural network to effectively solve the problem of dark channel prior algorithm in color distortion of sky region in foggy situations. On the other hand, an optimized dark channel prior defogging algorithm is adopted to deal with the non-sky areas, which improves the brightness while preserving the excellent defogging performance of the dark channel prior algorithm. The simulation result indicates the superior performance of our approach comparing to the results provided by two recent researches in the area. The approach has outstanding performance of defogging both sky and non-sky regions, while the restored image has a great level of color reduction and without losing details. The algorithm also perfectly satisfies the performance indices of the image quality such as entropy information, visibility ratio, average gradient, and the saturation percentage, which is a good indication of the excellent performance of this model.